# Modulation Instability and Pattern Formation in Spatially Incoherent Light Beams

**Detlef Kip,**[(1,2)] **Marin Soljacic,**[(1,3)] **Mordechai Segev,**[(1,4)]
**Evgenia Eugenieva,**[(5)] **and Demetrios N. Christodoulides**[(5)]

[(1)] Physics Department, Technion, Haifa 32000, Israel
[(2)] Physics Department, Universität Osnabrück, 49069 Osnabrück, Germany
[(3)] Physics Department, Princeton University, Princeton, NJ 08544, USA
[(4)] Department of Electrical Engineering, Princeton University, Princeton, NJ 08544, USA
[(5)] Electrical Engineering and Computer Science Dept., Lehigh University, Bethlehem, PA 18015, USA

Modulation Instability (MI) is a universal process that appears in most nonlinear wave systems in nature. Because of MI, small amplitude and phase perturbations (from noise) grow rapidly under the combined effects of nonlinearity and diffraction (or dispersion, in the temporal domain). As a result, a broad optical beam (or a quasi-CW pulse) tends to disintegrate during propagation [1,2], leading to filamentation [2] or to break-up into pulse trains [1]. In general, MI typically occurs in the same parameter region where another universal phenomenon, soliton occurrence, is observed. Solitons are stationary localized wave-packets (wave-packets that never broaden) that share many features with real particles. For example, their total energy and momentum is conserved even when they interact with one another [3]. Intuitively, solitons can be understood as a result of the balance between the broadening tendency of diffraction (or dispersion) and nonlinear self-focusing. A soliton forms when the localized wave-packet induces (via the nonlinearity) a potential and "captures" itself in it, thus becoming a bound state in its own induced potential. In the spatial domain of optics, a spatial soliton forms when a very narrow optical beam induces (through self-focusing) a waveguide structure and guides itself in its own induced waveguide. The relation between MI and solitons is best manifested in the fact that the filaments (or the pulse trains) that emerge from the MI process are actually trains of almost ideal solitons.

Therefore, MI can be considered to be a precursor to soliton formation. Over the years, MI has been systematically investigated in connection with numerous nonlinear processes. Yet, it was always believed that MI is inherently a coherent process and thus it can only appear in nonlinear systems with a perfect degree of spatial/temporal coherence. Earlier this year however, our group was able to show theoretically [4] that MI can also exist in relation with partially-incoherent wave-packets or beams. This in turn leads to several important new features: for example, incoherent MI appears only if the 'strength' of the nonlinearity exceeds a well-defined threshold that depends on the degree of spatial correlation (coherence). Moreover, by appropriately suppressing MI, new families of solitons are possible which have no counterpart whatsoever in the coherent regime [5]. This prediction of incoherent MI actually reflects on many other nonlinear systems beyond optics: it implies that patterns can form spontaneously (from noise) in nonlinear many-body systems involving weakly-correlated particles, such as, for example, electrons in semiconductors at the vicinity of the quantum Hall regime, high-$T_c$ superconductors, and atomic gases at temperatures slightly higher than Bose-Einstein-Condensation (BEC) temperatures. **Here, we present the first experimental observation of modulation instability of partially spatially incoherent light beams in non-instantaneous nonlinear media. We show that even in such a nonlinear partially coherent system (of weakly-correlated particles) patterns can form spontaneously. Incoherent MI occurs above a specific threshold that depends on the beams' coherence properties (correlation distance), and leads to a periodic train of one-dimensional (1D) filaments. At a higher value of nonlinearity, incoherent MI displays a two-dimensional (2D) instability and leads to self-ordered arrays of light spots.**

Before we proceed to describe incoherent MI, we revisit the main ideas that make incoherent solitons happen. Until a few years ago, solitons were considered to be solely coherent entities. However, in 1996 the first experimental observations of solitons made of partially-spatially-incoherent light [6] and in 1997 of temporally and spatially incoherent ("white") light [7] have proven beyond doubt that incoherent solitons do exist. This sequence of events has opened up entirely new directions in soliton science. Since then, numerous theoretical and experimental papers have been published on bright and dark incoherent solitons: their range of existence and their structure [8-11], their interactions [12], their stability properties [13], and their relation to multimode composite solitons [14]. The existence of incoherent solitons proves that self-focusing is possible not only for coherent wave-packets but also for wave-packets upon which the phase is random. The key to their existence is the non-instantaneous nature of the nonlinearity, which responds only to the beam's time-averaged intensity structure and not to the instantaneous highly speckled and fragmented wave-front. In other words, the response time of the nonlinear medium must be much larger than the average time of phase fluctuations across the beam. Thus, the time-averaged intensity induces, through the nonlinearity, a multimode waveguide structure (a potential well that can bind many states), whose guided modes are populated by the optical field with its instantaneous speckled structure. With this non-instantaneous nature of the nonlinearity in mind, **we were motivated to find out if patterns can form spontaneously on a partially coherent uniform beam, through the interplay between nonlinearity and diffraction, in a such a random-phase wave front of uniform intensity.** As a first step, we have shown theoretically [4] that a uniform yet partially-incoherent wave-front is unstable in such nonlinear media, provided that the nonlinearity exceeds a well-defined threshold that is set by the coherence properties. Above that threshold, MI should occur, and patterns should form. Here, we experimentally verify these predictions, and reveal a series of new features that relate to the appearance of 2D ordered lattices of filaments.

The main predictions of the incoherent MI theory [4] are as follows. (I) The existence of a sharp threshold for the nonlinear index change, below which perturbations (noise) on top of a uniform input beam decay and above which a quasi-periodic pattern forms. (II) The threshold depends upon the coherence properties of the input beam: the threshold increases with decreasing correlation distance (decreasing spatial coherence). (III) Saturation alone, while keeping the maximum index change and correlation distance fixed, arrests the growth rate of the MI and can decrease it to below the MI threshold. In the next sections we describe our experimental results that confirm all of these predictions and also reveal new unexpected features.

In our incoherent MI experiments, we use a strontium-barium niobate (SBN:60) crystal and employ its photorefractive screening nonlinearity [15,16], which is of a saturable nature. The dimensions of the sample are $a \times b \times c = 7.0 \times 6.5 \times 8.0 mm^3$, where light propagation is along the crystalline a-axis and the external electric bias field is applied along the c-axis. At moderate intensities (1 Watt/$cm^2$) the response time of our crystal is $\tau \sim 0.1$ seconds, thus for any light beam across which the phase varies much faster than $\tau$, the nonlinear crystal responds only to the time-averaged (over times much larger than $\tau$) intensity structure. In our experimental setup, we split a cw argon ion laser beam (of $\lambda=514.5$ nm wavelength) into two beams using a polarizing beamsplitter. Each beam is sent through a rotating diffuser which introduces a random phase varying much faster than $\tau$, acting as a source of partially spatially incoherent light. Following the rotating diffusers the beams are expanded, collimated and made uniform, and recombined using another polarizing beamsplitter. Finally both beams are launched into the crystal and co-propagate in it. When an external (bias) dc field is applied to the crystal, the extraordinarily-polarized beam experiences a large index change, and thus serves as the "signal beam", whereas the ordinarily-polarized beam experiences only a tiny index change and therefore serves as a background beam (its only role is to tune the degree of saturation of the nonlinearity [16]). A lens and a polarizer are used to image the signal beam intensity at the output face of the sample onto a CCD camera. We control the degree of coherence of the signal beam by adjusting the diameter of the laser beam incident on the rotating diffuser: the larger the beam diameter, the higher the incoherence and the shorter the correlation distance $l_c$. The background beam is made highly incoherent, which guarantees that it never forms any patterns. We estimate the correlation distance $l_c$ (at the input face of the crystal, when the system is linear: zero applied field) as the average value of the full-width-half-maximum (FWHM) of the speckle size on the CCD camera when the diffuser is momentarily stopped.

Upon the application of a sufficiently large bias field to the crystal, the signal beam experiences MI and forms patterns (Fig. 1). When the input signal beam is uniform, the underlying nonlinearity is of the form

$$\delta n = \Delta n_0 [1+ (I_0 / I_{sat})] [ I(r) / (I(r) + I_{sat} ) ] , \qquad (1)$$

where $I(r)$ is the local intensity as a function of coordinate r, $I_{sat}$ is the intensity of the incident background beam, and $I_0$ is the intensity of the signal beam at the input face. The term $[1+ (I_0 / I_{sat})]$ comes from the fact that the total current flowing through the crystal is practically the photocurrent generated by both beams (in contradistinction with the case of bright screening solitons, where the soliton beam is very narrow compared to the crystal width and thus does not affect the photocurrent, and this factor is equal to unity [15]). In Eq. 1, $\Delta n_0=-0.5n_e^3 r_{33}(V/L)$ is the electrooptic refractive index change, $n_e$ is the extraordinary refractive index, $r_{33}$ is the electrooptic tensor element, and (V/L) is the externally applied electric field.

Incoherent MI is observed for a nonlinearity δn exceeding a certain threshold. When an external voltage is applied to the nonlinear crystal with a magnitude large enough to allow for MI, the homogeneous light distribution at the output face of the sample becomes periodically modulated and starts to form 1D filaments of incoherent light. In our experiments, the preferred direction of the stripes is perpendicular to the c-axis of the crystal. We believe that this is due to the existence of striations in our sample, which act as 'initial noise' that is eventually amplified by MI. These are index inhomogeneities in planes perpendicular to the c-axis that originate from melt composition changes during growth of the crystal. Another possible reason for the preferential 1D directionality might have to do with the anisotropy of the photorefractive nonlinearity. However, **the final orientation of the stripes is rather random**, with the largest observed angle of the inclination of the stripes relative to the c-axis being roughly 45 degrees. Figure 1 shows typical examples of MI of partially spatially incoherent light. Shown is the intensity of the signal beam in the output plane of the nonlinear crystal. The coherence length of the incoherent light is $l_c$=17.5 μm and the intensity ratio $I_o/I_{sat}$=1. Figure 1(a) shows the output intensity without nonlinearity (V/L=0). The cases (b), (c), and (d) correspond to a value of the nonlinearity just below the threshold for 1D incoherent MI, at threshold, and just above the threshold. This shows beyond any doubt (1) the existence of incoherent MI, and (2) that incoherent MI occurs only when the nonlinear index change exceeds a well-defined threshold. In particular, Fig. 1(c) shows a mixed state exactly at threshold, in which order and disorder coexist. This is a clear indication that the nonlinear interaction undergoes an order-disorder phase transition. These phenomena were predicted in our theoretical paper [4]. But the experiments, as often happens in science, reveal new surprises. When the nonlinearity is further increased, a second threshold is reached: the filaments become unstable [Fig. 1(e)] and start to break into **an ordered array of spots (2D filaments)** as shown in Fig. 1(f). We emphasize that in all the pictures displayed in this figure, the correlation distance is much shorter than the distance between two adjacent stripes or filaments. Thus, this work is a clear demonstration that pattern can form also in weakly-correlated nonlinear multi-particle systems.

Next, we study the dependence of the MI threshold on the coherence properties of the beam. For a constant intensity ratio $I_0/I_{sat}$, the threshold where MI occurs depends on the incoherence of the light and on $\Delta n_0$ (which we control through the applied voltage V). To identify the MI threshold experimentally, one needs to examine the growth dynamics of perturbations and observe whether they grow or decay. Obviously, this is very hard to measure, especially because the initial perturbations originate from random noise. Instead, we investigate the visibility (modulation depth) of the pattern observed at the output face of the crystal: random fluctuations that do not increase have a tiny (less than 5%) visibility, whereas the perturbations that grow emerge at high visibility (>50%) stripes. We have conducted numerous experiments with various degrees of coherence of the input beam, and measure the modulation depth of the output stripes as a function of the applied field (which is translated to $\Delta n_0$). Figure 2(a) displays the results: it shows the modulation depth $m = (I_{max}-I_{min})/(I_{max}+I_{min})$ of the light at the output plane, as a function of $\Delta n_0$ for different correlation distances $l_c$ and $I_o/I_{sat}=1$. For the case of a fully coherent input beam, m becomes large even at a vanishingly small nonlinearity. This is because coherent MI has no threshold. When the correlation distance is reduced, however, a well-defined threshold is observed: the jump from very low visibility to a large visibility is always abrupt, because for every beam with a finite $l_c$ there is always a threshold for MI. Clearly, the MI threshold shifts towards higher value of $\Delta n_0$ with decreasing correlation distance $l_c$.

Once the nonlinearity exceeds the threshold for incoherent MI, the transverse frequencies that have gain grow exponentially and form the periodic patterns shown in Fig. 1. This growth leads to a large modulation depth (high visibility) in the output patterns, and, equally important, to a considerable deviation of these stripes from a pure sinusoidal shape, i.e., the propagation dynamics becomes highly nonlinear. Part of this dynamics was captured in the last figure in [4], by the appearance of the second spatial harmonic, yet the experiment provides considerably more insight into the nonlinear dynamical evolution of the patterns. Figure 2(b) shows typical intensity cross-sections of the stripes at the output plane. In this particular set of data, $l_c=17.5$ µm and $\Delta n_o$ values of $2.75*10^{-4}$ (i), $4.0*10^{-4}$ (ii), $5.0*10^{-4}$ (iii), and $8.0*10^{-4}$ (iv). At the lowest $\Delta n_o$ value, MI is barely above threshold (i). For the higher value at (ii), the modulation depth is higher yet the stripes have a sinusoidal shape. At the high value of (iii), the shape of the stripes is no longer sinusoidal and several higher harmonics participate. For an even higher nonlinearity, the spectrum becomes irregular (iv), and 2D breakup into filaments starts to occur.

The periodicity (or the spatial frequency) of the 1D filaments that emerge in the MI process depends on the coherence properties of the beam and on the magnitude of the nonlinearity. In a saturable nonlinear medium, for a given intensity ratio, the spatial frequency should monotonically increase with increasing correlation distance and with increasing $\Delta n_o$ [4]. We indeed observe this trend in our experiments. Figure 3(a) shows the dominating spatial frequency $f_{max}$ (number of stripes per unit length) of the output intensity as a function of correlation distance $l_c$ for a constant intensity ratio $I_o/I_{sat}=1$ and three different values of $\Delta n_o$. The 1D incoherent MI theory, when applied to the nonlinearity of Eq. (1), with the parameters $\lambda=514.5$ nm, $n_e=2.3$, $r_{33}=260$ pm/V, and assuming a Lorentzian-shaped angular power spectrum of the incoherent light, results in the plots shown in Fig. 3(b). Clearly, there is a good qualitative agreement between theory and experiments. The 1D incoherent MI theory also predicts the dependence of the dominating spatial frequency $f_{max}$ on $\Delta n_o$, for a given correlation distance, as shown in Fig. 3(d). In this case, the experiments confirm the increase of $f_{max}$ with increasing $\Delta n_o$, and shift towards higher $\Delta n_o$ values that occurs for decreasing $l_c$. However, experimentally, we observe (Fig. 3(c)) a turning point in the plots: the spatial frequency reaches a clear maximum and then goes down for increasing $\Delta n_o$. This is a new feature that was not predicted by the 1D theory. In fact, for $\Delta n_o$ values larger than those of the peaks, the 1D stripes become irregular and start to break up into 2D filaments, and this leads to a decrease in the spatial frequency of the stripes. These new unexpected effects should be addressed in the context of a 2D incoherent MI theory. It is very possible that a 2D linearized MI analysis (similar to that of [4] but of a higher dimension) will not suffice to explain the breakup into an array of filaments, and will have to involve either heavy computation or a perturbative nonlinear treatment.

Up to this point, the nonlinearity in our experiments had the form given in Eq. (1), which is not saturable. Based on the 1D incoherent MI theory [4], we expect that saturation of the optical nonlinearity should arrest the MI growth rate. To investigate saturation effects, we have modified the nature of our photorefractive screening nonlinearity in a rather easy way: by launching a "flat top" beam that is narrower than the distance between the electrodes in our crystal, yet at the same time is wide enough to serve as a "quasi-uniform beam" at its flat-top. Since the beam is finite, it does not contribute to the total current flowing through the crystal at steady state. Hence, the nonlinearity is now $\delta n = \Delta n_0 [ I(r) / (I(r) + I_{sat}) ]$, which is the more commonly-used form of the photorefractive screening nonlinearity [15], and it has a **saturable** nature. When we launch such a beam in a biased crystal with $\Delta n_0 = 6*10^{-4}$, and with a ratio between the peak intensity and the saturation intensity, $I(0)/I_{sat} = 3$, patterns form in several regions on the beam, as shown in Fig. 4. At the flat-top of the beam, low visibility stripes appear. In this region the nonlinearity is above threshold but in rather deep saturation, so the MI growth rate is suppressed. Then, at the margins of the beam, where the local ratio $I(r) + I_{sat}$ is around and slightly below unity, high-visibility stripes appear. In this region the nonlinearity is above threshold and at the same time it is not saturated, so the MI growth rate is large. Finally, at the far margins of the beam, the local nonlinearity is below threshold, because $I(r) << I_{sat}$. A nice by-product of this particular experiment is the clear evidence (in Fig. 4) that the 1D stripes emerge at different orientations, and are not affected much by the local noise (striations). We expect that similar experiments with incoherent MI in the saturated regime and high nonlinearities that lead to 2D "lattices" of filaments (as shown in Fig. 1), will reveal a wealth of new features, because the "lattice" will now form features of varying order and varying scales in different regions of the beam. This seems to be a fascinating option and we are currently investigating it in our laboratory.

Before closing, we would like to relate our nonlinear optical system to other nonlinear systems of weakly-correlated particles. Our prediction and experimental observation implies that in **all such systems, patterns will form spontaneously, provided that the nonlinearity is larger than a threshold value, which in turn is set by the correlation distance.** For example, we expect that 1D and 2D patterns will form in an atomic gas at temperatures slightly above the Bose-Einstein-Condensation temperature: at temperatures at which the atoms possess independent degrees of freedom (and cannot be described by a single wave function, as in the BEC state), yet they are still weakly correlated. At least for atoms with attractive collision forces (a negative scattering length), such patterns should form, depending on the dimensionality of the problem describing the atoms in the trap. The equation governing the evolution of the "mean field" of an atomic gas is the so-called Gross-Pitaevski equation [17], which is literally identical to the nonlinear wave equation in nonlinear optics that gives rise to (1+1)D Kerr solitons. ***However, to the best of our knowledge, no such ideas of pattern formation through modulation instability have been previously suggested previously for atomic gases [18].*** Yet in other areas of physics, in fact, there are already at least some hints that such patterns do exist in disordered many-body nonlinear systems. To be specific, several experimental papers have reported a large anisotropy in the resistivity of a two-dimensional electron system with weak disorder [19]. The observed anisotropy is now attributed to the combination of nonlinear transport and weak disorder [20], which is exactly the transport-equivalent of an optical nonlinearity and incoherence in optical systems such as ours. The theoretical works predict the existence of 1D stripes (electron stripes) of charge density waves [20]. Spontaneous formation of stripes was also predicted and observed in high-$T_c$ superconductors [21], which is again a nonlinear weakly-correlated many-body system. Finally, as we have already discussed in Ref. [4], spontaneously-forming patterns are also known in at least one system of classical particles: in a gravitational system. The spontaneous emergence of patterns in all of these diverse fields of science indicates that pattern formation in nonlinear weakly-correlated systems is a universal property. It is a gift of nature that in optics we can study it directly, visualizing every little detail of the physics involved, and at the same time, being able to isolate the underlying effects and develop a theory that captures the core effects that are indeed universal. Our hope is to be able to understand such patterns in a higher dimensionality, as we already observe them in our laboratory, and to draw new exciting implications to other fields of nature.

In conclusion, we have presented the first experimental observation of modulation instability of spatially-incoherent light beams in a nonlinear optical system. We have proven that modulation instability occurs when the nonlinearity exceeds a sharp threshold. The MI threshold depends on the degree of incoherence, that is, the correlation distance, or the maximum distance between two points upon the beam that are still phase-correlated. We have shown that during the MI process a homogeneous input wave-front breaks into one-dimensional stripes. For a higher nonlinearity, we observe a second threshold at which the stripes become unstable and form (spontaneously) a spatially ordered pattern of two-dimensional filaments (spots).

This work was supported by the Israeli Science Foundation, NSF, ARO, AFOSR, and the Deutsche Forschungsgemeinschaft.

**Figures**

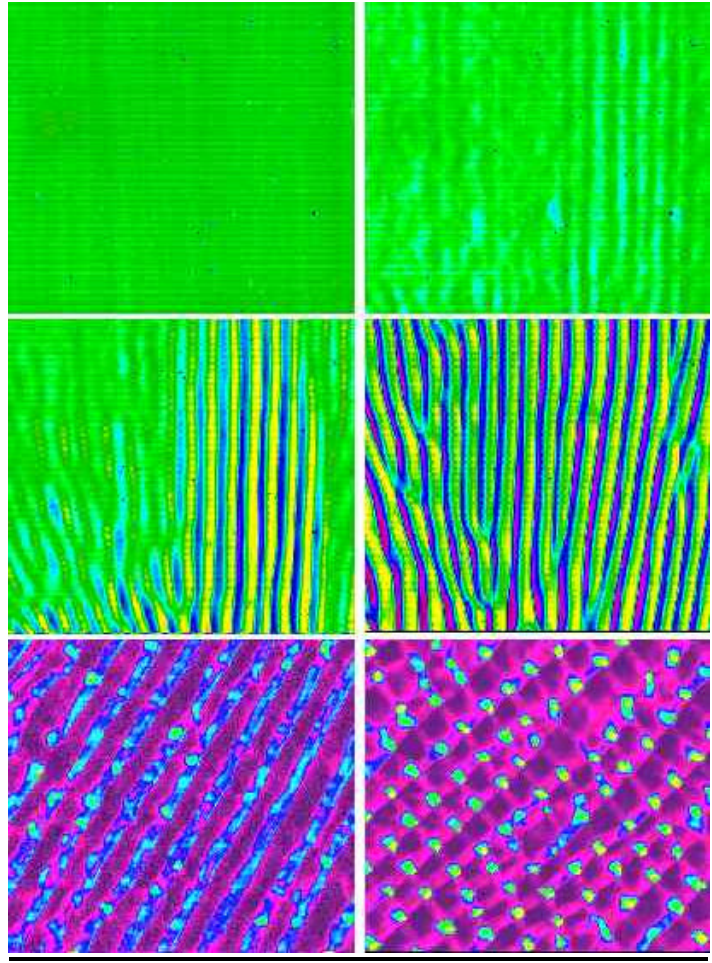

**Fig. 1.** The intensity structure of a partially-spatially-incoherent beam at the output plane of the nonlinear crystal. The sample is illuminated homogeneously with partially spatially incoherent light with a coherence length $l_c=17.5$ μm. The displayed area is $1.0\times1.0$ mm$^2$ (a-d) and $0.8\times0.8$ mm$^2$ (e,f), respectively. The size of the nonlinear refractive index change of the crystal is successively increased from (a) $\Delta n_0=0$ (the linear case), to (b) $3.5*10^{-4}$, (c) $4.0*10^{-4}$, (d) $4.5*10^{-4}$, (e) $9*10^{-4}$ (e), and (f) $1*10^{-3}$. The plots (b-d) show the cases just below threshold (no features), at threshold (partial features), and just above threshold (features everywhere) for 1D incoherent MI that leads to 1D filaments. Far above this threshold, at a much higher value of the nonlinearity, the 1D filaments become unstable (e), and finally become ordered in a regular two-dimensional pattern (f).

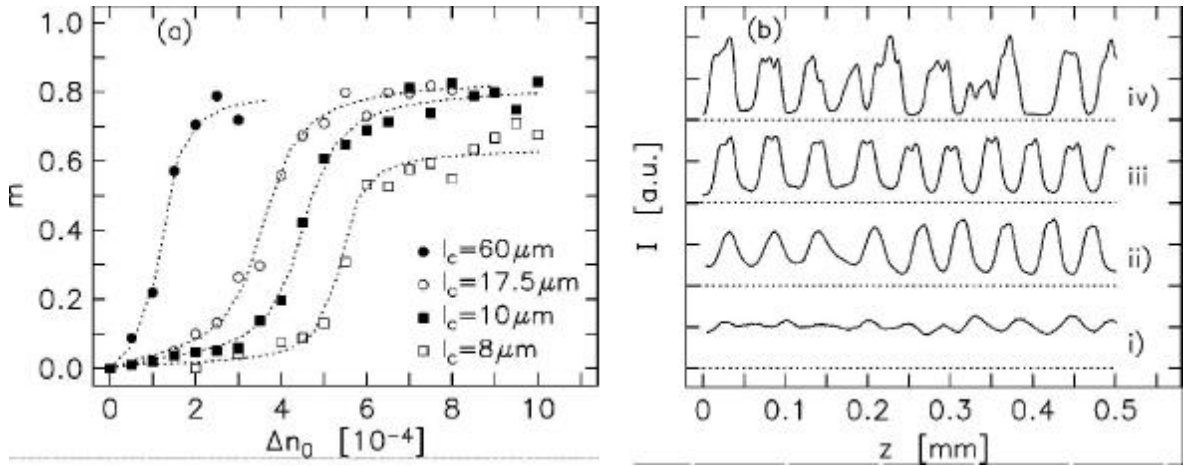

**Fig. 2.** Threshold dependence of incoherent MI: modulation $m=(I_{max}-I_{min})/(I_{max}+I_{min})$ of the light pattern vs. size of the nonlinearity $\Delta n_o$ for different correlation distances $l_c$ and an intensity ratio $I_o/I_{sat}=1$. a) measured values of m for $l_c$=8, 10, and 17.5 µm and for coherent light ($l_c \to \infty$). The dotted curves are guides for the eye. b) intensity cross-sections of the stripes for $l_c$=17.5 µm and a nonlinear refractive index change of $\Delta n_o=2.75*10^{-4}$ (i), $4.0*10^{-4}$ (ii), $5.0*10^{-4}$ (iii), and $8.0*10^{-4}$ (iv). The dotted lines indicate the base line of the respective profile. The stripes emerge as sinusoidal stripes (for nonlinearity just above threshold), and turn into square-wave stripes at a higher nonlinearity, and eventually break up into filaments at a large enough nonlinearity.

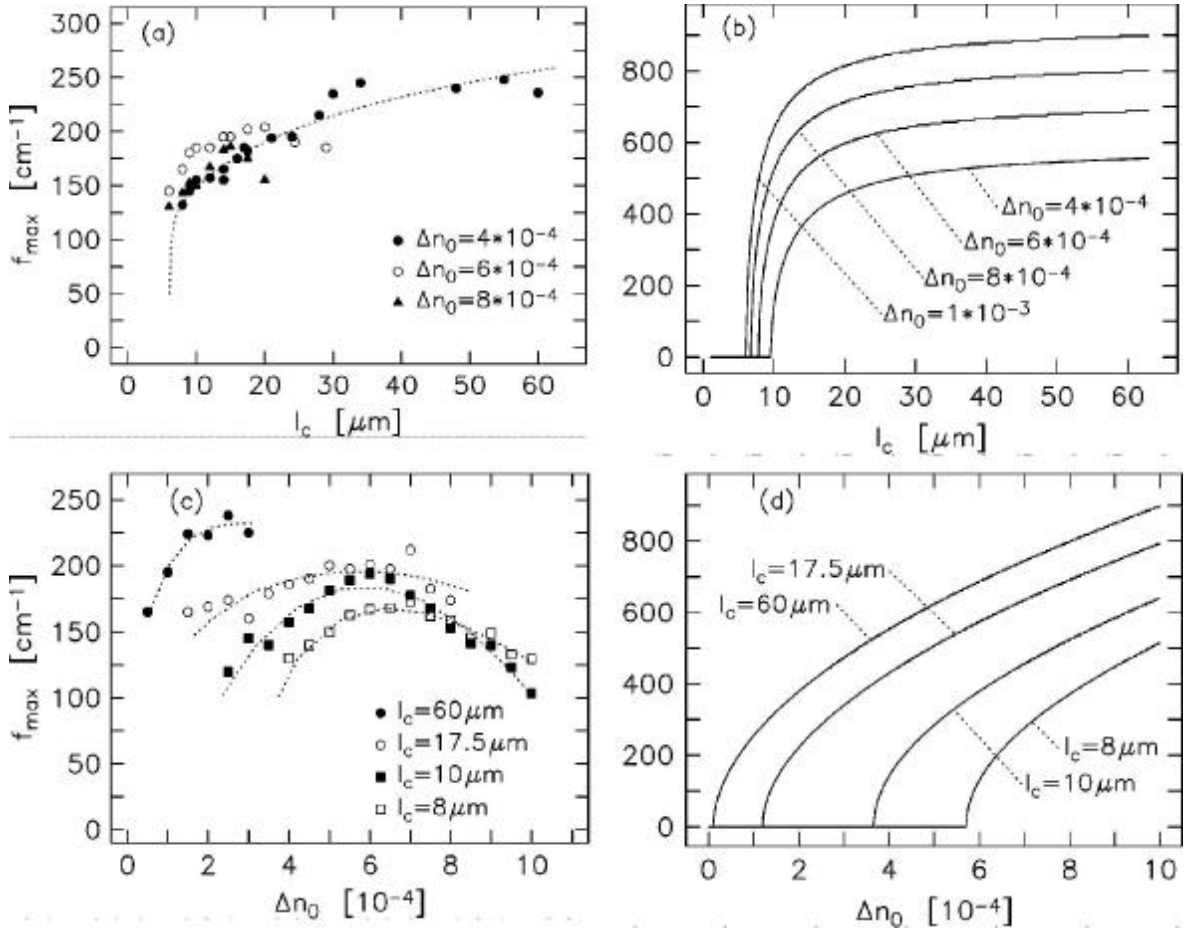

**Fig. 3.** The dominating spatial frequency $f_{max}$ (number of stripes per unit length) of the output intensity as a function of correlation distance $l_c$ (a, experiment; b, theory) and nonlinear refractive index change (c, experiment; d, theory). All measured spatial frequencies are for an experimental parameter range where the 1D filaments are stable and 2D instability is not yet visible. The theoretical curves in (b) and (d) are deduced form the 1D model described in [4] using the nonlinearity of Eq. 1 and $\lambda=514.5$ nm, $n_e=2.3$, and $r_{33}=260$ pm/V. The dotted curves in (a) and (c) are guides for the eye.

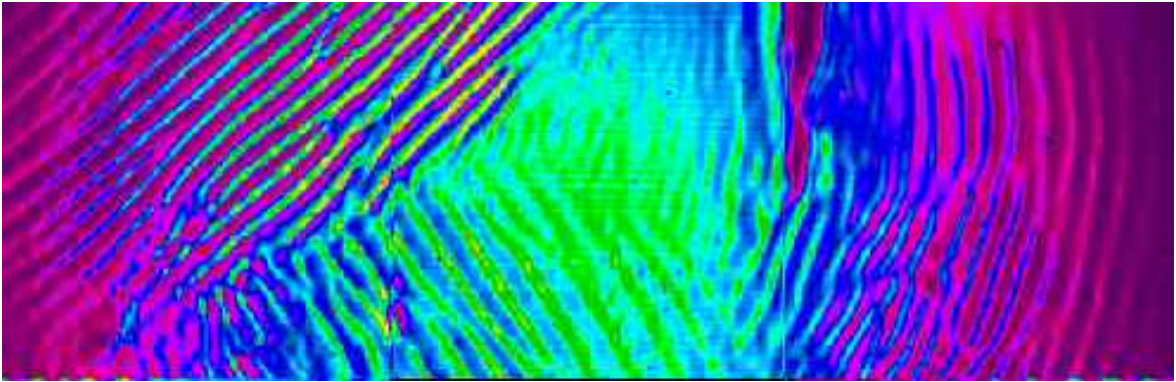

**Fig. 4.** Suppression of incoherent MI due to saturation of the nonlinearity. The intensity structure of a finite signal beam (gaussian beam with a width (FWHM) of 1 mm) at the output plane of the crystal. The intensity ratio (peak of beam to background/saturation intensity) is $I_0/I_{sat}=3$. Without nonlinearity ($\Delta n_0=0$), the output beam shows no features. The photograph is taken for $\Delta n_0=6*10^{-4}$. The saturation nature of the nonlinearity clearly suppresses MI in the center of the beam, whereas strong modulation and filaments of random orientation occur in the margins of the beam.